\documentclass[11pt]{article}
\pdfoutput=1
\usepackage{jheppub}
\usepackage{subfig}
\usepackage{comment}
\usepackage{array,setspace,mathrsfs,amsfonts,amsmath,yfonts,dsfont,bbm,colonequals}
\usepackage{graphicx} 
\usepackage{amsmath}
\usepackage{tikz}
\usepackage{amssymb}
\usepackage{amsfonts}
\usepackage{appendix}
\usepackage{comment}
\usepackage{color}
\usepackage{amsmath}
\usepackage{xcolor}
\usepackage{tikz}
\usepackage{amsthm}
\usetikzlibrary{decorations.pathmorphing}

\title{Seeing double: shock waves and the de Sitter horizon}
\author{Willy Fischler*, Hare Krishna*, and Sarah Racz$^{\dagger}$}

\affiliation{
$*$ Weinberg Institute, Department of Physics, University of Texas at Austin, Austin, TX 78712, USA\\
$\dagger$ Department of Physics, Sarah Lawrence College, Bronxville NY 10708, USA\\}

\emailAdd{fischler@physics.utexas.edu}\emailAdd{hkrishna.phy@gmail.com}\emailAdd{sracz@sarahlawrence.edu}

\begin{document}
\hfill       UT-WI-38-2025
\abstract{We consider a de Sitter observer in his rest frame at late times who observes a particle slightly displaced from unstable equilibrium. Initially, the observer notices an axisymmetric and parity-violating deformation along the trajectory of the displaced particle of his cosmological horizon. On a time scale of order $\ell$, the de Sitter radius, the particle is nearly absorbed by the cosmological horizon and has been accelerated to an ultra-relativistic speed and thus is well approximated as a shock wave. In the shock wave limit, the observer sees an axisymmetric deformation of his horizon with parity restored, which we interpret as arising due to a particle from the complementary static patch. We comment on the holographic implications of this result and note that there is no need to extend the holographic screen of de Sitter spacetime beyond the empty static patch to account for this signal.}

\maketitle

    \section{Introduction and motivation}

    For three decades now, the holographic principle has provided powerful insights into the nature of quantum gravity \cite{susskind_world_1995, hooft_dimensional_2009}. Though there are many examples of holographic dualities, none have been explored more than the AdS/CFT correspondence \cite{maldacena_large-n_1999}. However, evidence suggests that our universe is accelerating with an equation of state consistent with a positive cosmological constant, which is well-described by de Sitter spacetime. Although there have been recent attempts to understand holography in more general spacetimes, a clear consensus has yet to emerge. What is clear is that if we wish to understand quantum gravity in our own universe, a quantum theory of de Sitter spacetime is necessary. \\

    In this work, we take the approach of `static patch holography’, which posits that the holographic screen of de Sitter spacetime lives on the cosmological horizon of the empty static patch. Various proposals, including the holographic space-time (HST) approach of Banks and Fischler \cite{Banks:2001px,hep-th/0609062} and the double-scaled SYK (DSSYK) proposal of Susskind \cite{2209.09999}, Narovlansky, and Verlinde \cite{Narovlansky:2023lfz}, are of this class of holographic theory. Such proposals use the insight of Gibbons and Hawking that de Sitter spacetime has entropy given by the Bekenstein-Hawking entropy law \cite{PhysRevD.15.2738} \begin{equation}S_{dS} = \frac{\mathcal{A}_{CH}}{4}\end{equation} for the cosmological horizon whose size is given by the de Sitter length scale $\ell$. Further `data' for a quantum description of de Sitter spacetime can be gleaned from the presence of a mass (black hole) in the spacetime. The Schwarzschild-de Sitter solution (SdS), which describes a static black hole within the de Sitter static patch, shrinks the radius of the cosmological horizon, thus reducing the entropy. It has been interpreted that empty de Sitter spacetime must be the maximally entropic state \cite{hep-th/0102077}. Further, the mass of a black hole in de Sitter spacetime has an upper bound given by the Nariai mass \begin{equation} M_{crit} = \frac{\ell}{3\sqrt 3}, \end{equation} which provides an ultraviolet cutoff on the `energy' in the static patch. There is an infrared cutoff provided by the size of the horizon. Moreover, since the maximal entropy of de Sitter spacetime is finite, these arguments led to the proposal that the Hilbert space of de Sitter spacetime is finite dimensional \cite{Fischler2000,Banks2000,Bousso:2000nf}. 
    
    In a series of recent papers \cite{Fischler:2024cgm,Fischler:2024idi}, the authors have sought to provide more data that a quantum theory of de Sitter spacetime should reproduce. By considering static, albeit unstable, configurations of masses, it was shown that the cosmological horizon encodes all of the information associated with extended objects placed within the de Sitter bulk. Small masses were arranged on the vertices of Platonic solids deep in the bulk and it was found that the cosmological horizon deformed to the dual of the bulk polyhedron. Though the area of the cosmological horizon under such deformations did not change, the sizes of bumps and dips on the cosmological horizon encoded the size of the bulk objects. The results are expected to hold for more general mass configurations and, in principle, allow for the reconstruction of the bulk configuration using data on the horizon. Introducing charge to the bulk configurations amplified the horizon deformations as the horizon was shown to be sensitive to the energy stored in the electric field associated with the bulk masses. The horizon was also shown to polarize and inherit an induced charge or rotation due to bulk charge or angular momentum, as required to satisfy Gauss's law. Though the bulk polyhedra previously considered are unstable to perturbations, we note that a single Schwarzschild de Sitter black hole is similarly unstable -- the consequences of which condition we study in this paper.
    
    In this work, we extend our analyses of the cosmological horizon beyond the static regime to the time-dependent case. Due to cosmic repulsion, any object in de Sitter spacetime perturbed away from the origin will fly to the horizon. The particle accelerates and approaches the speed of light as it closes in on the cosmological horizon in a time $t\sim O(\ell)$. Initially, when the particle is slightly displaced from its equilibrium location deep in the bulk, it travels at non-relativistic velocities. We use a post-Newtonian expansion around the Schwarzschild-de Sitter solution to study the effect of the moving particle on the cosmological horizon. At late times, we model the particle as a shock wave living on the horizon for small shock wave energies $p\ll \ell$. 

    To see the cosmological horizon, an observer sitting at $r=0$ must wait an infinite amount of time, which washes out any time-dependent effects. We account for what an observer can see at observation time $T_o$ by looking at the observer's backwards light cone, which we dub the \textbf{observer horizon}. In our calculations, we consider late time observers ($T_o$ large), so the observer horizon is of the same order as the cosmological horizon.

    We show that the observer horizon is sensitive to the moving mass both when the mass is initially displaced and moves at small velocities and after it has accelerated to the speed of light. As expected, the effect of the moving mass introduces time-dependent axisymmetric deformations of the observer horizon. We find that in the small velocity limit the horizon deformation breaks parity along the direction of the particle's motion, taken to be the $z$ axis. We find that the horizon dips towards the moving particle, which is compensated by an antipodal protrusion. Surprisingly, in the shock wave limit we find parity restored along the $z$ axis despite having performed a parity-violating boost along $+z$. The presence of a second `spike' on the horizon suggests that a localized de Sitter observer has access to information from the complementary static patch. It is well known that shock waves lead to a time advance in de Sitter spacetime , which puts two opposing static patches in causal contact and makes the spacetime's Penrose diagram ``taller'' \cite{Gao:2001ut,Leblond:2002ns}. Crucially, though one observer has information about both patches, the original empty de Sitter horizon can encode all that data.

    Throughout this paper we work in Planck units. The rest of the paper is organized as follows. In section \ref{preliminaries}, we review some basics of de Sitter geometry and show how the shock wave metric is obtained from a boost of the Schwarzschild de Sitter metric. In section \ref{observerhorizon}, we define the observer horizon and calculate the effect of non-stationary masses on the horizon. We conclude in section \ref{discussion}, with some discussion on the holographic implications of our work. Detailed calculations are presented in the appendices. 

    \section{de Sitter preliminaries}
    \label{preliminaries}
    \subsection{Nothing is stable in the static patch}
    In the weak-field limit of general relativity, a particle in the de Sitter static patch is subject to the Newtonian potential  of an inverse harmonic oscillator $\Phi_{dS}=- \frac{r^2}{\ell^2} $. Thus any particle in the static patch is unstable under a small perturbation and will approach the speed of light as it accelerates towards the cosmological horizon. For a short while after a particle deep in the bulk has been displaced, it travels with a small, nearly constant velocity. In the infinite-boost limit, after $t\sim\ell$, the spacetime describing the accelerated particle can be described as the empty de Sitter spacetime with a shock wave perturbation. The shock wave spacetime has been studied extensively in the Minkowski spacetime by Aichelburg and Sexl \cite{1971GReGr...2..303A} and Dray and t'Hooft \cite{Dray:1984ha}. The shockwave geometry has been generalized to curved spaces and, in particular, to de Sitter spacetime \cite{Hotta:1992qy,Sfetsos:1994xa,McLoughlin:2025shj,Bittermann:2022hhy,PhysRevD.56.4756,Aalsma:2021kle}. In this work, we will follow mostly the convention of Hotta-Tanaka \cite{Hotta:1992qy}, who obtained the de Sitter shockwave geometry by performing a boost of the small-mass Schwarzschild de Sitter spacetime. Although the shock wave is an exact solution to the Einstein equation, we will treat it as a perturbation over the empty de Sitter solution when we calculating the observer's horizon in section \ref{observerhorizon}.

    We now review some of the preliminaries of de Sitter embeddings before turning our attention to cosmological shock wave solutions.

    \subsection{de Sitter geometry}
    \label{background geometry}
    In four dimensions, empty de Sitter spacetime can be understood as a hyperboloid embedded in a five-dimensional Minkowski space, described by coordinates $Z_{i }$, where $i= 0,\ldots, 4$, with the constraint
    \begin{equation}\ell^2 = -Z_0^2 + Z_1^2 + Z_2^2 + Z_3^2 + Z_4^2, \end{equation}
    where $\ell$ is the de Sitter radius.
    The region of de Sitter space accessible to a single observer can be obtained through a transformation to the `static' coordinate system given by
    \begin{equation}
    Z_0 = \sqrt{\ell^2 - r^2}\sinh(t/\ell), \quad Z_4 = \pm \sqrt{\ell^2 - r^2}\cosh(t/\ell), \quad Z_i = rz_i,
    \end{equation}
    where $1\leq i \leq 3$ and $z_i$ are the coordinates on a sphere.
    \begin{eqnarray}
        z_1= r \cos \theta, \quad z_2= r \sin \theta \cos \phi,\quad  z_3= r \sin \theta \sin \phi.
    \end{eqnarray}

\begin{figure}
        \centering
        \includegraphics[width=0.75\linewidth]{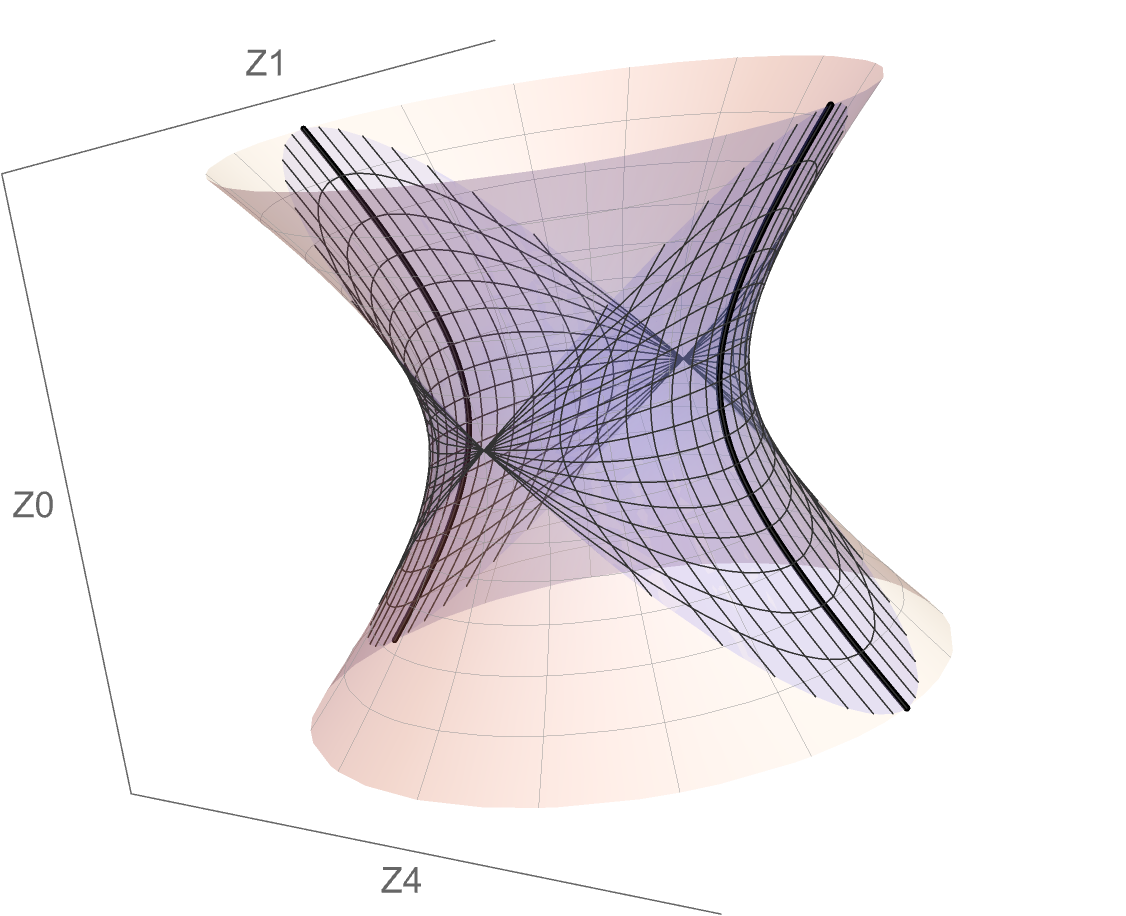}
        \caption{The de Sitter hyperboloid embedding is shown with the $Z^2$ and $Z^3$ coordinates suppressed. The timelike world lines of two test particles in opposing static patches are shown in black as perturbations living on the empty de Sitter background.}
        \label{fig:dsembed}
    \end{figure}

    In these coordinates the metric is given by
    \begin{equation}
        ds^2_{dS} = -\left( 1- \frac{r^2}{\ell^2}\right) dt^2 + \left( 1- \frac{r^2}{\ell^2}\right)^{-1} dr^2 + r^2 d\Omega_2^2,
    \label{eqn:static patch metric}
    \end{equation}
    where $d\Omega$ is the metric on the 2-sphere. This metric admits a timelike Killing vector, making it manifestly time-translation invariant. An observer at $r=0$ is surrounded by a cosmological horizon located at $r_{\mathcal H} = \ell$. The two signs of $Z_4$ lead to complementary static patches (right and left), which are out of causal contact with one another. The right and left static patches on the global de Sitter hyperboloid can be seen in Fig. \ref{fig:dsembed}. The world lines of test particles placed at $r=0$ in either patch are also shown explicitly. 

    
    \subsection{Schwarzschild de Sitter}
    
    The static patch metric \ref{eqn:static patch metric} can be generalized to include a spherically symmetric mass at $r=0$. Such solutions are given by the Schwarzschild de Sitter (SdS) spacetime, which describes a black hole inside the static patch. The SdS metric is given by
    \begin{equation}
        ds^2_{SdS} = -\left( 1- \frac{2m}{r} -\frac{r^2}{\ell^2}\right) dt^2 + \left( 1- \frac{2m}{r} -\frac{r^2}{\ell^2}\right)^{-1} dr^2 + r^2 d\Omega_2^2,
    \end{equation}
    where $m$ is the mass of the black hole. This metric admits two horizons, the cosmological horizon $r_{\mathcal H}$ and the black hole horizon $r_\mathcal{B}$. The presence of a black hole at the center of one static patch necessitates a second black hole in the opposing patch. This can be understood as the requirement that the Hamiltonian constraint be satisfied on a spatial slice of global de Sitter.

    Throughout our calculations, we take the black hole mass $m$ to be small, as otherwise the $d=5$ Minkowski hyperboloid embedding described above would be invalid. In this limit, we treat the black hole as a perturbation on the empty de Sitter background. To leading order in $m$, the Schwarzschild de Sitter metric is given by 
    \begin{equation}
        ds^2_{SdS} = -\left( 1- \frac{2m}{r} -\frac{r^2}{\ell^2}\right) dt^2 + \left( 1 -\frac{r^2}{\ell^2}\right)^{-1} \left[1 +\left(1-\frac{r^2}{\ell^2}\right)^{-1} \frac{2m}{r}\right] dr^2 + r^2 d\Omega_2^2.
    \end{equation}
    It is easy to show that the cosmological horizon is located at $r_{\mathcal H}=\ell-m$ and that the black hole horizon is located at $r_{\mathcal{B}}= 2 m$.
    
    Expressed in the embedding coordinates $Z_i$, the linearized metric takes the form 
    \begin{eqnarray}
     ds_{SdS}^2 &&= ds_{dS}^2 + \frac{2 m \, \ell^2}{(Z_4^2-Z_0^2)^2\sqrt{\ell^2+Z_0^2-Z_4^2}}[(\ell^2Z_4^2+\ell^2 Z_0^2+Z_0^2Z_4^2-Z_4^4)d Z_0^2\nonumber\\
        &&-2 (2 \ell^2+Z_0^2-Z_4^2)Z_0\,Z_4 \,dZ_0\, dZ_4+(\ell^2 Z_4^2+\ell^2 Z_0^2+Z_0^4-Z_0^2Z_4^2)dZ_4^2],
        \label{SdSembed}
    \end{eqnarray}
    where $ds_{dS}^2$ is the empty static patch metric. 
    
    \subsection{Cosmological shock wave solutions}
\begin{figure}
        \centering
        \includegraphics[width=0.75\linewidth]{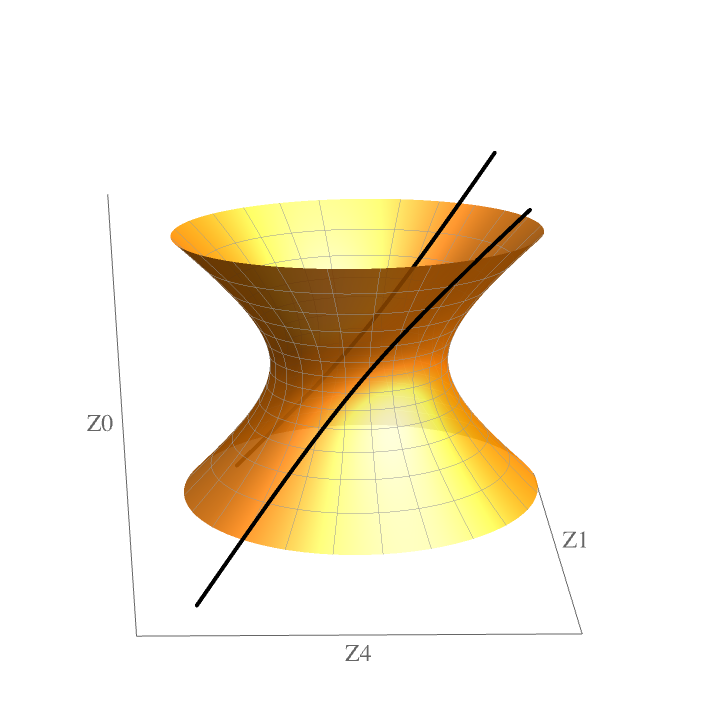}
        \caption{The de Sitter hyperboloid embedding is shown with the $Z^2$ and $Z^3$ coordinates suppressed. The world lines of two particles in opposing static patches that have been boosted are shown in black as perturbations living on the empty de Sitter background.}
        \label{fig:shockwaveembed}
    \end{figure}

    We start with the linearized Schwarzschild de Sitter solution, and boost the mass along the $Z_1$ direction with a boost parameter $v$ ($0\leq v\leq 1$)

    \begin{equation}
    Z_0 = \frac{Z_0' + v Z_1'}{\sqrt{1 - v^2}}, \quad Z_1 = \frac{Z_1' + v Z_0}{\sqrt{1 - v^2}}, \quad Z_2' = Z_2, \quad Z_3' = Z_3, \quad Z_4' = Z_4. 
    \end{equation}
    To find the shock wave metric, we take the ultra-relativistic limit $v \to 1$ and use the distributional identity
    \begin{equation}
    \lim_{v \to 1} \frac{1}{\sqrt{1-v^2}} f\Bigg(\frac{(Z_0+vZ_1)^2}{1-v^2}\Bigg)=\delta(Z_0+Z_1)\int_{-\infty}^{\infty} f(x^2) dx
    \end{equation}
    to rewrite \eqref{SdSembed} as 
    \begin{equation}
    \label{shockwavemet}
    ds^2 = ds_{\text{dS}}^2 + ds_{sh}^2(p),
    \end{equation}
    where $ds_{\text{dS}}^2$ is the empty de Sitter metric, and
    \begin{equation}
    ds_{sh}^2(p) = 4\, \frac{p}{\ell} \delta(Z_0 + Z_1) \left[ -2\ell + Z_4 \ln \frac{\ell + Z_4}{\ell - Z_4} \right] (dZ_0 + dZ_1)^2
    \end{equation}
    describes the shockwave. The energy of the particle is defined to be $p = \frac{m}{\sqrt{1 - v^2}}$, so we send $m\to0$ in order to keep $p$ finite.  In appendix \ref{shockenergy}, we map this parameter $p$ to the initial position and velocity. In Fig. \ref{fig:shockwaveembed}, we show how these particles in both patches are boosted in the embedding coordinates. If we instead boost the particle to $v\to 1-\epsilon$ and work to leading order in $\epsilon$ we find a broadening of the delta function which we show in Appendix \ref{delta function}.

    For our purposes, we consider an observer located in the right static patch which we describe with  Kruskal-Szekeres-like coordinates for de Sitter spacetime 
    \begin{eqnarray}
    \label{statickruskal}
           u= -\ell \sqrt{\frac{\ell-r}{\ell+r}} e^{-t/\ell},\quad    v= \ell \sqrt{\frac{\ell-r}{\ell+r}} e^{t/\ell}.
       \end{eqnarray}
    where $u<0,v>0$. The left patch coordinates can be obtained by sending the $t \rightarrow t-i\pi\ell$, which effectively reverses the sign in the above relation.

    In these coordinates, the shockwave metric \eqref{shockwavemet} becomes

    \begin{eqnarray}
    \label{shock wave metric}
        ds^2=2 A(u,v)(-du dv)- 2 A(u,v) \delta(v) f(\theta) dv^2+g(u,v) d\Omega^2, 
    \end{eqnarray}
    where
    
    \begin{eqnarray}
        A= \frac{2 \ell^4}{(uv-\ell^2)^2},\quad g= \Big(\frac{u v+\ell^2}{u v-\ell^2}\Big)^2 \ell^2,\quad f(\theta)= p\Big(2- \cos \theta \log \frac{1+\cos \theta }{1-\cos \theta}\Big).
    \end{eqnarray}
    The presence of the $\delta$ function, a coordinate artifact, informs us that the shock is located on the $v=0$ plane. Following Dray and t'Hooft \cite{Dray:1984ha}, the spacetime is described by empty de Sitter spacetime for $v<0$, but for $v>0$ the spacetime has a shockwave. We shift the coordinates to account for the shock and define 
    \begin{eqnarray}
    \label{eqn:hat transform}
        \hat{v}= v, \quad \hat{u}= u+ \Theta[v] f, \quad \hat{\theta}= \theta,\quad  \hat{\phi}=\phi,
    \end{eqnarray}
    where $\Theta[v]$ is the step function.
    In the hatted coordinates, the shock metric becomes
    \begin{eqnarray}
      ds^2= -2 A(\hat{v},\hat{u}-\Theta[\hat{v}] f) d\hat{v} (d\hat{u}- \Theta[\hat{v}] \partial_{\theta} f \, d\theta )+ g(\hat{v},\hat{u} -\Theta[\hat{v}] f) d\Omega^2.
    \end{eqnarray}

\begin{figure}
        \centering
        \includegraphics[width=\linewidth]{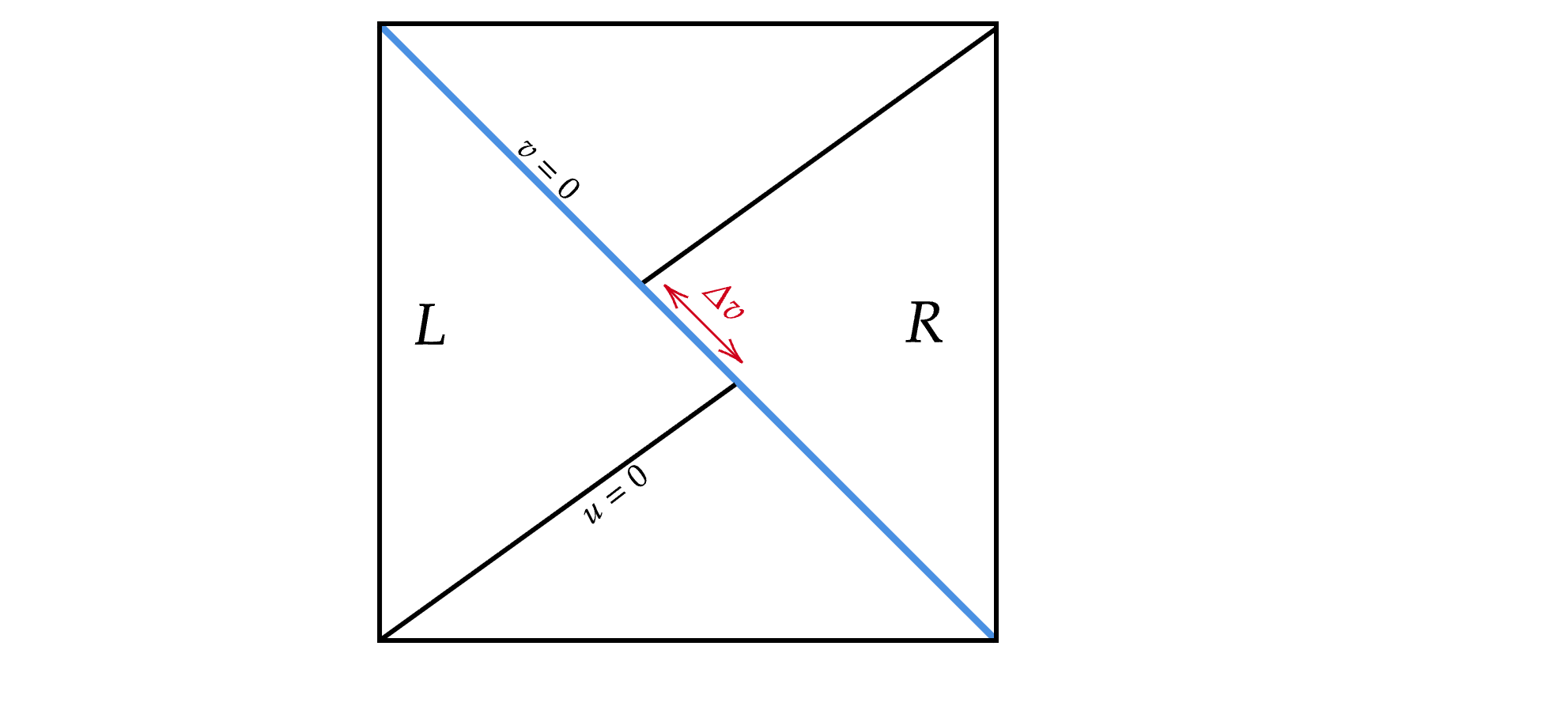}
        \caption{In this Penrose diagram, a shock wave (shown in red) is along the $u$ axis. The observer is in the right patch. Due to the shock wave, the $ u$ coordinate is shifted by $\Delta u=-\Theta (v) f(\theta)$. Effectively, the shock makes the diagram taller, and a signal from the left patch may come in to causal contact with the observer in the right patch.}
        \label{fig:shockwaveshift}
    \end{figure}

    Geodesics who cross the past horizon, $v=0$, encounter the shock and experience a time advance in the $u$ coordinate. This is shown in Fig. \ref{fig:shockwaveshift}. In effect, this allows for the right and left static patches to come into causal contact with one another \cite{Leblond:2002ns,Aalsma:2021kle}.

    The shock wave is an exact solution to Einstein's equations with stress tensor
    \begin{equation}
    \begin{aligned}
     T_{vv}&=  \frac{p}{2 \pi \ell^2}\Big[ \delta(\cos \theta-1)+\delta(\cos \theta+1)\Big]\delta(v),
     \end{aligned}
    \end{equation}
    which arises due to the singular boost transformation.
    
    \section{Changes to the horizon}
    \label{observerhorizon}
    \subsection{Observer Horizon}
    We now define the observer horizon to be what an observer sitting at $r=0$ sees after waiting a time $T_o$. This is simply the backward light cone of the observer, which limits to the cosmological horizon at $T_o \to \infty$. Such a construction is necessary to see the effect of transient (time of $O(\ell)$) changes to the horizon. To compute the observer horizon, we consider a photon leaving the observer horizon, $r=r_h$, at an emission time  $t=t_i$ that reaches the observer at $r=0$ at the observation time $t=T_o$. The location of $r_h$ defines the observer horizon, and without loss of generality, we set $t_i=0$. Perturbations on empty de Sitter spacetime will leave imprints on the observer horizon (in its shape and size) that an observer would detect in finite time, though they would wash out as $T_o \to \infty$.\\
    \subsection{Small boost limit}
    \label{small boost}
We first consider the case of a black hole shortly after it has been perturbed away from the center of the static patch. Deep in the bulk, space is nearly flat, so the force of cosmic repulsion is negligible. In this limit, the black hole moves at a small, $v \ll 1$, constant velocity along the $+z$ axis. We study the effect of the black hole's motion on the observer horizon for a time interval $\tau\sim m <<T_o, \ell$. Measured by the observer's clock, the particle starts moving at time $t= T_o- \tau$, where $T_o$ is the observation time of a photon emitted from the observer horizon. The trajectory of particle is
\begin{eqnarray}
    r(t)= v \, t- v\, (T_o-\tau).
\end{eqnarray}
The Newtonian potential due to a moving particle in flat space is
\begin{eqnarray}
    \Phi= -\frac{ m}{|r- r(t) \cos \theta|}=-\frac{ m}{|r- (v t- v (T_o-\tau))\cos \theta|},
\end{eqnarray}
which we insert into the Schwarzschild-de Sitter metric by taking the coordinate transformation $r\rightarrow|r- r(t) \cos \theta|, t\rightarrow t$ in the Schwarzschild term. We work with a post Newtonian approximation to linear order in $v$, and treat any $v$-dependent term as a perturbation on a Schwarzschild-de Sitter background. 
The non-zero metric components are 
\begin{eqnarray}
\label{timedepe}
 &&   g_{tt}=-\left(1-\frac{2 m}{|r- r(t) \cos \theta|}-\frac{r^2}{\ell^2}\right), \quad g_{rr}= \frac{1}{\left(1-\frac{2 m}{|r- r(t) \cos \theta|}-\frac{r^2}{\ell^2}\right)},\\ &&g_{tr}= - 2v \cos \theta ,\quad      g_{\theta \theta}= r^2, \quad g_{\phi\phi}=r^2 \sin^2 \theta.
\end{eqnarray}
 A light ray traveling from the horizon is subject to the Schwarzschild-de Sitter background for times $0\leq t < T_0-\tau$ and feels the effect of the moving particle from $T_o-\tau\leq t \leq t_0$. To find the observer horizon we compute the geodesic along this light ray. 

\subsubsection{First part of the trajectory: $t \in [0, T_o-\tau]$} 
Initially, the photon travels in the Schwarzchild de-Sitter spacetime. The null geodesic, $ds^2=0$,  can be solved using the method of partial fractions. We write the integral as $I_1$
   \begin{eqnarray}
   I_1=\int_{0}^{T_o-\tau} dt=  -\int_{r_h}^{r_*} \frac{dr}{(1-\frac{2 m   }{r}-\frac{r^2}{\ell^2})},
    \end{eqnarray}
where $r_*=r|_{T_0-\tau}= 2 \sqrt{ m \tau}+ \frac{2}{3}v \cos \theta\,\, \tau $ is the radial position of the photon at $t=T_o-\tau$. \\

\subsubsection{Second part of the trajectory: $t \in [ T_o-\tau,T_o]$} 
After $t = T_o - \tau$, the light ray's geodesic is modified due to the motion of the mass. Since the mass's motion introduces a $g_{tr}$ component to the metric the null radial geodesic is given by
\begin{eqnarray}
\label{null1}
 g_{tt} dt^2+g_{rr} dr^2+2 g_{tr} dt \, dr=0. 
\end{eqnarray}
Solving the quadratic equation for $dr/dt$ gives
\begin{eqnarray}
    \frac{dr}{dt}= \frac{-\sqrt{g^2_{tr}-g_{rr}
   g_{tt}}-g_{tr}}{g_{rr}}=-\frac{2 m [r+v \cos (\theta ) (\tau
   -T_o+t)]}{r^2}.
\end{eqnarray}
Since we work with small velocity, $v <<1$, we solve the geodesic equation perturbatively. We use the ansatz $r(t)= r_0(t)+v r_1(t)$. We integrate from $r=r_*$ at $t=T_o - \tau$ to $r=0$ at $t=T_o$ and find
\begin{eqnarray}
    r(t)= 2 \sqrt{ m (T_o-t)+v^2 \tau^2 \cos^2 \theta/4}+ v\, \tau \cos \theta+ v \cos \theta\, \times\left(\frac{t- T_o}{3}\right)+O[v^2].
\end{eqnarray}
This is the radial distance light travels in the $[T_o-\tau, T_o]$ interval. \\

The total path the photon takes is the sum of the two parts of the trajectory. Schematically, this is given by
\begin{eqnarray}
    T_o&&= I_1|_{r_h}^{r_*}+ 2\sqrt{m \tau} +2 v \cos \theta \tau/3\\ \nonumber
    &&=\frac{m r_h^2}{\ell^2-r_h^2}-m \log\Big[\frac{
   (\ell^2-r_h^2)4 m^2}{r_h^2\ell^2}\Big]
+\frac{1}{2} \ell \log
   \left(\frac{\ell+r_h}{\ell-r_h}\right)+\frac{2
   v \tau \cos \theta}{3}.
\end{eqnarray}

We invert the above relation to find the location of the observer horizon $r_h$ as a function of the observer time $T_o$ 
\begin{eqnarray}
\label{small_boost_horizon}
    r_h=&& \ell- m- 2 e^{-2 T_o/\ell}\ell +4 m e^{-\frac{2 T_o}{\ell}} \\ \nonumber &&+4 m e^{-\frac{2 T_o}{\ell}} \log \left(\frac{16 m^2}{\ell^2} e^{-\frac{2 T_o}{\ell}}\right)-\frac{8}{3}v \tau \cos \theta e^{-\frac{2 T_o}{\ell}}.
\end{eqnarray}
This equation is valid for long observation times ($T_o$ large) and as $T_o \to \infty$ the observer horizon reduces to the (small mass) Schwarzschild de Sitter cosmological horizon. 

\begin{figure}
    \centering
    \includegraphics[width=0.5\linewidth]{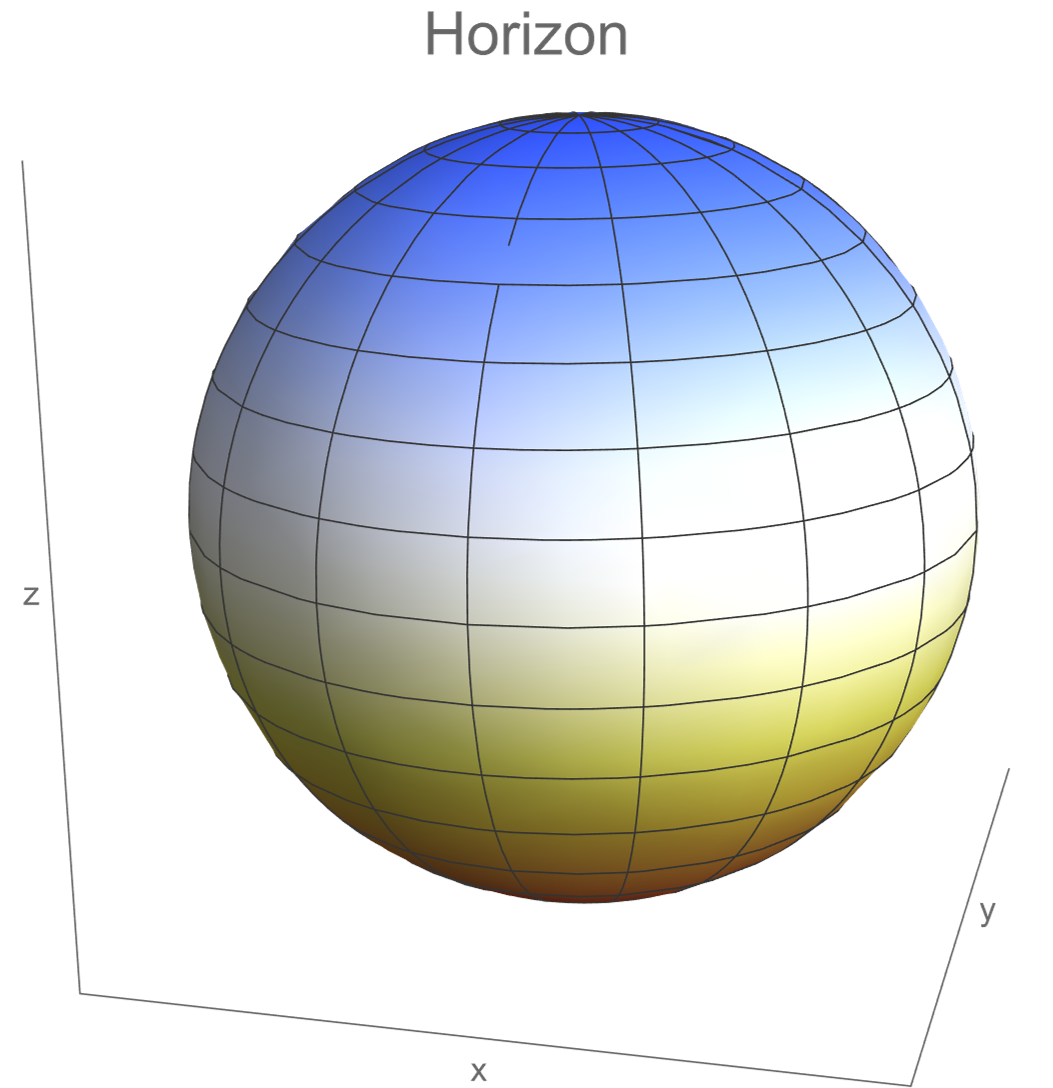}
    \caption{This is the horizon as seen by the observer when the particle is moving slowly and governed by eq \eqref{small_boost_horizon}. The blue region near the North Pole shows the dip, while the yellowish region near the South Pole depicts the protrusion. In this case, the particle is moving along the $+ z$ axis and breaks parity. We plot the horizon using parameters $\ell=1000, \, m=100,\, T_o=2000,\,\tau=1,\, v=0.1$, which are outside the range of perturbative validity, to qualitatively illustrate the horizon's shape.}
    \label{fig:small boost}
\end{figure}

The shape of the cosmological horizon is shown in the figure \ref{fig:small boost}. We observe that the horizon dips in towards the moving particle and bulges out at the opposite pole, thereby breaking parity. The horizon encodes the velocity of the moving particle, meaning that data is accessible from horizon measurements.  

    \subsection{Shock wave limit}
\begin{figure}
        \centering
        \includegraphics[width=0.5\linewidth]{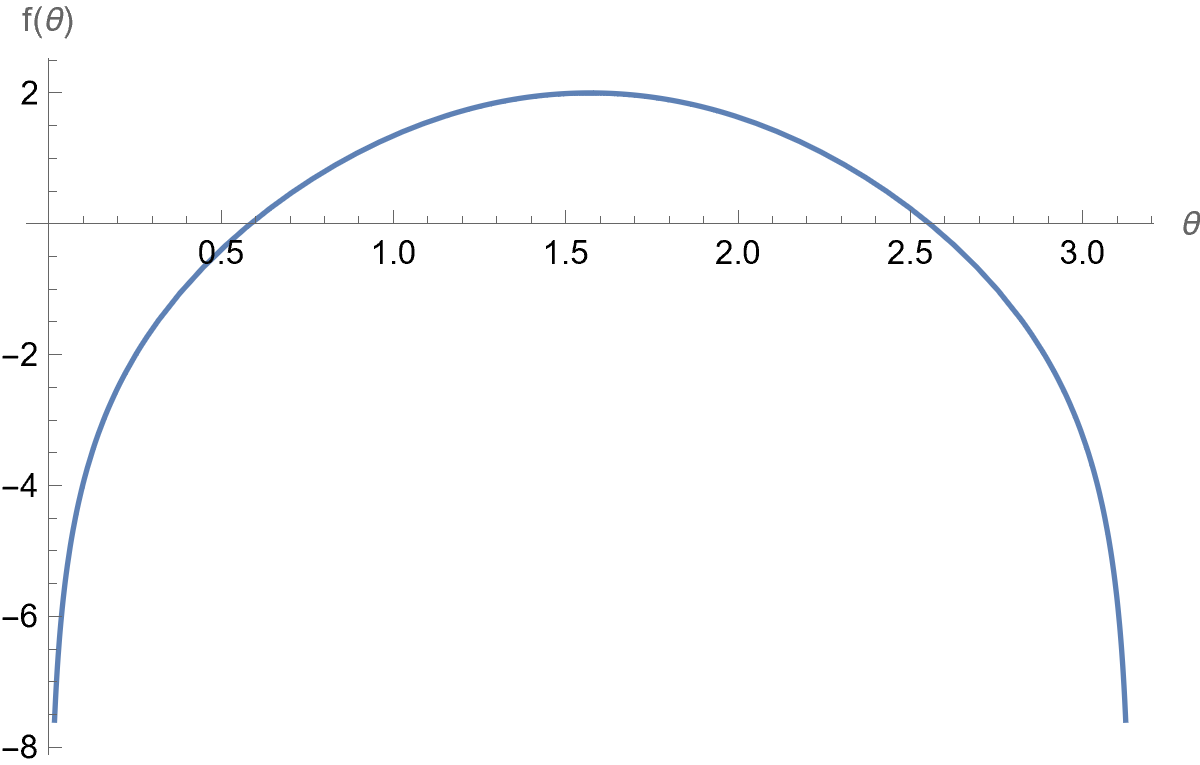}
        \caption{ Plot $f(\theta)$ against the angular coordinate $\theta$ with $p=1$. }
        \label{fig:ftheta}
    \end{figure}

    We now turn our attention to times $t\sim \ell$ after the particle has been displaced from equilibrium and has thus been accelerated to near-light speeds. The spacetime is well approximated by the cosmological shock wave solution whose metric can be found in \eqref{shock wave metric}. To track what an observer at $r=0$ can see, we analyze the geodesics that originate observer horizon $r_h$ at $t=0$ that reaches the observer at time $t=T_o$. In the null coordinate $u$, the light ray's initial and final conditions are given by 

        \begin{align}
            u_i&=-\ell \sqrt{\frac{\ell-r_h}{\ell+r_h}} \\ 
            u_f&= -\ell e^{-T_o/\ell}.
        \end{align}
    The null radial geodesic for $u$, which uses $v$ as an affine parameter, is \footnote{The null geodesic with $\dot{\theta}$ contribute at the order of $p^2$, and we work only in the linear order in $p$. This is analyzed carefully the appendix \ref{geodesic}.} given by
       \begin{eqnarray}
       -  \frac{du}{dv}-  f \delta (v) =0  
       \end{eqnarray}
    Using the transform \eqref{eqn:hat transform} to hatted coordinates, we incorporate the $\delta$ function into our $u$ coordinate and the geodesic equation becomes $$\frac{d\hat{u}}{dv}=0,$$ which implies $\hat{u}= \hat{u}_0.$ In the original, un-hatted coordinates we have $\Delta u=-\Theta (v) f(\theta)$. A light ray from the observer horizon will thus be shifted and experience a time advance when it crosses the shock. The shift is given by
    \begin{eqnarray}
        \Delta u= - f(\theta)=-\ell e^{-T_o/\ell}+\ell \sqrt{\frac{\ell-r_h}{\ell+r_h}}.
    \end{eqnarray}
    We solve the above expression for the observer horizon $r_h$ and find \footnote{ The function $f(\theta)$ is not always positive. It is plotted in the figure. \ref{fig:ftheta}. Hence the shift in $\Delta u$ is positive for angle $[0,33]$ and $[148,180]$ and negative in the range $[33,148]$.} 
      \begin{equation}\begin{aligned}
          r_h&=\ell-2 \ell\, e^{-\frac{2(T_o)
       }{\ell}}-4 \,f(\theta )\, e^{-T_o/\ell}\\
       &=\ell-2 \ell\, e^{-\frac{2T_o
       }{\ell}}-4 \,\,   e^{-T_o/\ell}p\Big(2- \cos \theta \log \frac{1+\cos \theta }{1-\cos \theta}\Big).
       \label{horizon deformed}
      \end{aligned}\end{equation}
      The observer horizon has an anisotropy (given by the time advance $f(\theta)$) which fades as more time elapses between the photon emission time and the time of observation $T_o$. We can see that as $T_o \to \infty$ we recover the area of the empty de Sitter cosmological horizon. 
      
      For small masses, it is well known that the de Sitter horizon shrinks to $r_\mathcal{H} = \ell - m$. Taking this seriously, as $m$ increases the horizon would change topology once $m\sim \ell$. There is a known upper bound on the mass of a black hole in de Sitter spacetime, the Nariai mass $M_{Nariai} = \frac{\ell}{3\sqrt 3}$. Using the same logic we expect there to be an upper bound on the shock wave energy $p \sim \ell$. 
      \begin{figure}
          \centering
          \includegraphics[width=0.35\linewidth]{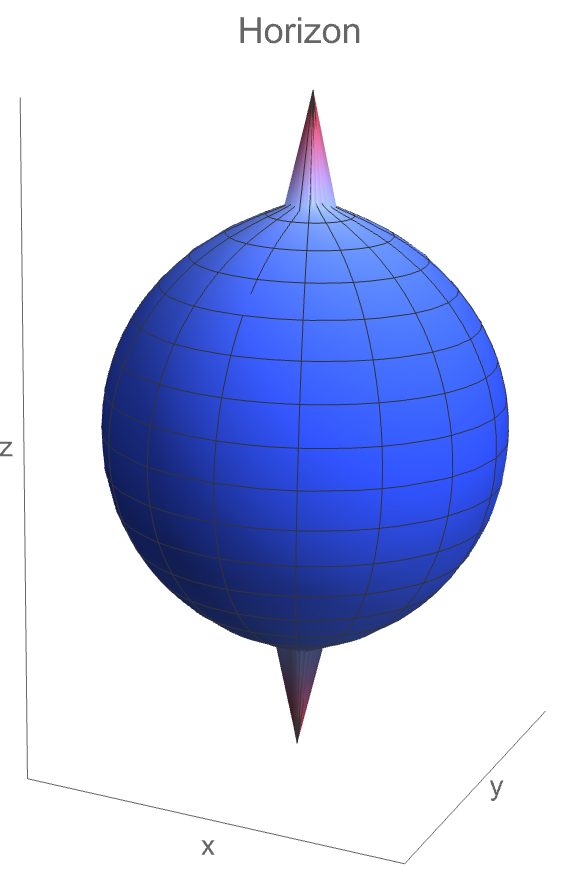}
          \caption{The observer horizon due to the shock wave has two bumps on the north and south poles of the sphere. This is the horizon governed by \eqref{horizon deformed} and shows that parity has been restored along the $z$ axis. In this plot, we use parameters $\ell=1000,\,  T_o=1000 ,\,  p=10$, which are outside the range of perturbative validity, to qualitatively show the horizon's shape.} 
          \label{hori}
      \end{figure}
    
    A plot of the observer horizon is shown in Fig. \ref{hori}. Surprisingly, the observer sees two spikes on the horizon at $\theta = 0$ and $\theta = \pi$ corresponding to the two poles in the logarithmic term of the anisotropy. We interpret the second spike as arising from the boosted particle from the complementary static patch. Since the two static patches are in causal contact due to the shock, information that was previously inaccessible can be detected by the right patch observer. 

    The horizon area is given by
      \begin{eqnarray}
          A_h&&= 2 \pi \int_{-1}^1 d(\cos \theta) [\ell^2-4 \ell^2\, e^{-\frac{2T_o
       }{\ell}}-8 \ell \,f(\theta )\,   e^{-T_o/\ell}]\nonumber\\
       &&= 2 \pi (2 \ell^2- 8 \ell^2  e^{-\frac{2T_o
       }{\ell}}- 8 \ell p\, e^{-T_o/\ell}),
    \end{eqnarray} 
    which is always smaller than the empty de Sitter horizon area. Holographically, the interpretation holds that empty de Sitter spacetime is the maximally entropic state and contains all of the information needed to describe what the right patch observer sees. We emphasize that the holographic screen of de Sitter spacetime does not need to be extended to account for the information originating in the left patch.

\section{Discussion}
\label{discussion}
In this work, we've studied observational signatures of non-stationary objects on the de Sitter cosmological horizon. To do this, we first introduced the \textit{observer horizon} as the backwards lightcone of an observer located at $r=0$ in the de Sitter static patch. Obviously, as the observation time $T_o \to \infty$ the observer horizon approaches the de Sitter empty cosmological horizon in the shockwave limit. In the slow-velocity limit the horizon limits to $r_\mathcal{H} = \ell - m$ since we worked around a Schwarzschild-de Sitter background.  The observer horizon allowed us to study time-dependent effects which would be washed out on the cosmological horizon.

We consider a particle that has been displaced from its equilibrium position deep in the de Sitter bulk and track its influence on the cosmological horizon. Initially, the perturbed particle feels a small force of cosmic repulsion, as space is nearly flat, and travels with a small and almost constant velocity. After $t\sim \ell$ the particle is ultra-relativistic after being accelerated towards the de Sitter horizon and is well approximated by a shock wave. We find that in both regimes an observer at $r=0$ will detect time-dependent changes to their observed horizon that encode the information of the moving mass. In the small velocity limit, we find the horizon inherits a parity-violating deformation along the direction of the particle's motion. The horizon dips towards the approaching mass and pulls away on the other side, leaving the horizon looking something like an onion. This horizon behavior aligns with the result that the cosmological horizon inherits the dual shape to the bulk matter configuration as seen in \cite{Fischler:2024cgm} and the fact that the presence of the mass in the static patch shrinks the cosmological horizon.

More interestingly, the observer horizon for the shock wave metric, which we use to model the particle after it has been accelerated to $v\sim1$, restores parity along the boost direction of the particle. As can be seen in Fig. \ref{hori}, the static patch observer sees two outward spikes along $\theta = 0$ and $\theta = \pi$ even though the black hole was only boosted along the $+z$ direction. We interpret the second spike as the signature of the particle, which was also boosted, from the complementary static patch. Due to the `tall diagram' effect, the observer in the right static patch has access to information from the complementary static patch which manifests on the cosmological horizon \cite{Gao:2001ut,Leblond:2002ns}. Though the observer seems to have access to an `extra' signal from the opposing static patch, the observer's empty cosmological horizon does not need to be extended to account for this information. The spikes on the cosmological horizon encode the shock wave energy $p$, the conserved charge associated with time translation invariance.

It is tempting to conclude that there is an upper bound on the shockwave energy $p$, where energy is defined to be the conserved charge associated with time translations. The shock wave energy can be written in terms of the particle's initial conditions, as can be seen in Appendix \ref{shockenergy}. In the Nariai limit, the spacetime is $dS_2 \times S^2$, which differs from the topology of Schwarzschild-de Sitter spacetime. The resulting horizon deformation from the shockwave \eqref{horizon deformed} hints towards a topology change for high enough values of $p$. Concluding an upper bound on $p$ is outside of our regime of perturbative validity though we do believe such a bound exists. Further, it would be instructive to find \textit{where} the transition between the parity violating and parity preserving horizon occurs.  
\\ \\ 
\subsection{Holographic implications}
Throughout this work, we have used the assumption that the horizon of empty de Sitter spacetime encodes all of the information accessible to the de Sitter observer. It is well known that the two complementary de Sitter patches are dynamically coupled in the presence of a shock wave \cite{Leblond:2002ns}, and we show that an observer can detect this effect from a measurement of his horizon and see information from `the other side'. As we've remarked, we find that the empty de Sitter horizon has enough information to encode both complementary static patches. In a private communication, Sekino and Susskind argue something along the same lines \cite{susskind}.\\ \\ 

\subsection{Gauss's law and induced stress tensor at the horizon} We offer an alternate interpretation for what the de Sitter observer detects arising from the membrane paradigm. De Sitter complementarity states that for a de Sitter observer, all physically accessible phenomena admit a description entirely within that observer's static patch, so instead of interpreting the second particle as originating from the complementary static patch, we interpret it as a source in the induced stress tensor on the cosmological horizon \cite{Parikh:1997ma}. 

Intuitively, this can be understood using electromagnetic charges. Suppose one places a positive charge in de Sitter spacetime. To satisfy Gauss' law, there must be a negative charge in the opposing static patch. However, the observer who only has information about his own static patch sees electric field lines end at the cosmological horizon. Thus, the cosmological horizon must have negative charge smeared over it so the net charge is zero. It was shown that the horizon inherits dual charges from the bulk in \cite{Fischler:2024idi}. In the gravitational case, a similar interpretation holds.

 \section*{Acknowledgements}
     We thank A.~Zimmermann and E.~C\'aceres for useful discussions. We thank L.~Susskind for sharing a draft of his forthcoming paper with Y.~Sekino. The work of H.K. is supported by NSF grant PHY-2210562 and CNS Spark Grant 2025-2029. The work of W.F. is supported by NSF grant PHY-2210562. S.R. was supported by the Sarah Lawrence Summer Science Program. S.R. thanks the hospitality of Centro de Ciencias de Benasque, where part of this work was completed during the conference ``Gravity - New quantum and string perspectives”.

      \appendix
      \section{Geodesic analysis:}
      \label{geodesic}
    
    The geodesic equation for the shock wave metric can be written as
    \begin{eqnarray}
       && \ddot{v}+\frac{\dot{\theta}^2 v \left(\ell^2+u v\right)+2
      \dot{v}^2 u+\dot{\phi}^2 v \sin ^2(\theta )
       \left(\ell^2+u v\right)}{\ell^2-u v}=0 \\
       && \ddot{\theta}+\frac{2  \delta (v) f'(\theta
       )}{\ell^2}\,\dot{v}^2+\frac{4 \,\dot{\theta}\,  l^2 ( \dot{u}
       v+\dot{v} u)}{\ell^4-u^2 v^2}-\dot{\phi}^2 \sin
       (\theta ) \cos (\theta )=0 \\
       &&\ddot{\phi}+2 \dot{\theta} \dot{\phi} \cot (\theta
       )+\frac{4 \dot{\phi} \ell^2 (\dot{u} v+\dot{v}
       u)}{\ell^4-u^2 v^2}=0\\
       &&\ddot{u}+\frac{1}{2} \dot{v}^2 \left(-\frac{4 u \delta (v)
       f(\theta )}{\ell^2}+2 f(\theta ) \delta '(v)\right)+2
       \dot{\theta} \dot{v} \delta (v) f'(\theta)\nonumber\\&&\quad\quad+\frac{\dot{\theta}^2 u \left(\ell^2+u v\right)+2
       \dot{u}^2 v+\dot{\phi}^2 u \sin ^2(\theta )
       \left(\ell^2+u v\right)}{\ell^2-u v}=0\nonumber\\
       \end{eqnarray}
     Here, the dot represents the derivative with respect to the affine parameter. We chose $v$ as an affine parameter. The energy corresponding to the null geodesic can be written as
       \begin{eqnarray}
       - 2A \frac{du}{dv}+ g \Big[(\frac{d\theta}{dv})^2+ \sin^2 \theta (\frac{d\phi}{dv})^2\Big]- 2 A f \delta (v) =0 
       \label{energy}
       \end{eqnarray}
    Now, $\dot{v}^2$ term with the delta function can be incorporated by a discontinuity in $v$ as
       \begin{eqnarray}
           \hat{u}= u+\theta (v) f(\theta).
       \end{eqnarray}
    Here, we are not solving the geodesic equation. We simply want to learn the change in the geodesic when it crosses the shock. Now we look at the equation for $\dot{\theta}$, which has a discontinuity as 
     \begin{eqnarray}
       \frac{d\theta}{dv}|{_{v=0^+}}-   \frac{d\theta}{dv}|_{{v=0^-}} =- 2 \frac{f'(\theta)}{\ell^2}|_{v=0}.
     \end{eqnarray}
     The $\dot{\phi}$ is continuous across the shock at $v=0$. Now we analyse the eq \eqref{energy}. After the redefinition of $u$ in terms of $\hat{u}$, the eq becomes

      \begin{eqnarray}
      - 2 A \frac{d \hat{u}}{dv}+ g (\frac{d\theta}{dv})^2=0
      \end{eqnarray}
      Now assuming $\frac{d\theta}{dv}|_{{v=0^-}}=0$, then the above equation give the change in $u$ as
      \begin{eqnarray}
          \Delta \hat{u}= c_1+\frac{2g}{A} \frac{ f'^2}{ \, \ell^4}=c_1+\frac{p^2 \left(2 \cot (\theta )+\sin (\theta )
       \log \left(\cot ^2\left(\frac{\theta
       }{2}\right)\right)\right)^2}{ \ell^2}
      \end{eqnarray}
       Here, $c_1$ is an order one constant, while the other term is order $p^2$, and we ignore the terms of the order $p^2$ in our analysis in section \ref{observerhorizon}.
    \section{Area element}
    In this appendix, we find the area element on the horizon, which is topologically a sphere.\\
    
     First, we consider a surface embedded in $\mathbb{R}^3$ described by spherical coordinates $(r,\theta,\phi)$ by
    \[
    r(\theta) = R(\theta) = r_0 - r_1 f(\theta).
    \]
    We will later restrict to our case where $r_0= \ell-2 \ell\, e^{-\frac{2(t_o)
       }{\ell}} $ and $r_1=4\,    e^{-t_o/\ell}$. First, we write the embedding coordinate as
    \[
    \mathbf{X}(\theta,\phi) = R(\theta)\,\hat r(\theta,\phi),
    \]
    where
    \[
    \hat r(\theta,\phi) = 
    (\sin\theta\cos\phi, \; \sin\theta\sin\phi, \; \cos\theta).
    \]
    We can find the tangent vectors as
    \[
    \partial_\theta \mathbf{X} = R_{,\theta}\,\hat r + R\,\hat \theta, 
    \qquad
    \partial_\phi \mathbf{X} = R \sin\theta \,\hat \phi,
    \]
    where $(\hat r,\hat\theta,\hat\phi)$ are the standard orthonormal spherical unit vectors.
    
    Since $R$ depends only on $\theta$,
    \[
    R_{,\theta} = \frac{dR}{d\theta} = -\,r_1 f'(\theta).
    \]
    Then the induced metric is
    \[
    g_{ab} = \partial_a \mathbf{X}\cdot \partial_b \mathbf{X}.
    \]
    The metric components can be written explicitly as
    \[
    \begin{aligned}
    g_{\theta\theta} &= R_{,\theta}^2 + R^2 
    = r_1^2 \big(f'(\theta)\big)^2 + \big(r_0 - r_1 f(\theta)\big)^2, \\
    g_{\theta\phi} &= 0, \\
    g_{\phi\phi} &= R^2 \sin^2\theta 
    = \big(r_0 - r_1 f(\theta)\big)^2 \sin^2\theta.
    \end{aligned}
    \]
    
    Thus, the line element is
    \[
    ds^2 = \Big[\,r_1^2 \big(f'(\theta)\big)^2 + (r_0 - r_1 f(\theta))^2\,\Big]\,d\theta^2
    + (r_0 - r_1 f(\theta))^2 \sin^2\theta \, d\phi^2.
    \]
    Now we write the term only linear in $r_1$ as
    \begin{eqnarray}
        ds^2 = \Big[ r_0^2 - 2r_0 r_1 f(\theta)\,\Big]\,d\theta^2
    + [r_0^2 - 2 r_0 r_1 f(\theta) ]\sin^2\theta \, d\phi^2.
    \end{eqnarray}
    For our case, for time time-dependent horizon, we have the induced metric as
     $r_0= \ell-2 \ell\, e^{-\frac{2(t_o)
       }{\ell}} $ and $r_1=4\,    e^{-t_o/\ell}$
    \begin{eqnarray}
       ds^2 = \Big[ \ell^2 - 8\,\ell \,\,     e^{-t_o/\ell} f(\theta)\,\Big]\,d\theta^2
    +\sin^2 \theta \Big[ \ell^2 - 8\,\ell \,\,     e^{-t_o/\ell} f(\theta)\,\Big] d\phi^2.  
    \end{eqnarray}
    
    \subsection*{Determinant and area element}
    The determinant of the induced metric is
    \[
    \det g = g_{\theta\theta} g_{\phi\phi} - g_{\theta\phi}^2
    = \Big(r_0^4- 2 r_0^3 r_1 f(\theta)\Big)=\Big(\ell^4- 8\,\ell^3 \,\,     e^{-t_o/\ell} f(\theta)\Big)\sin^2 \theta
    \]
    Then the area element can be written as
    \begin{eqnarray}
        \sqrt{g} d\theta d\phi= \ell^2 \sin \theta\Big[1- \frac{4}{\ell}  e^{-t_o/\ell} f(\theta)\Big] d\theta \, d\phi
    \end{eqnarray}
    
    \section{The delta function}
    \label{delta function}
    In the main text, we have used
    the distributional identity
    \begin{equation}
    \lim_{v \to 1} \frac{1}{\sqrt{1-v^2}} f\Bigg(\frac{(Z_0+vZ_1)^2}{1-v^2}\Bigg)=\delta(Z_0+Z_1)\int_{-\infty}^{\infty} f(x^2) dx.
    \end{equation}
    We want to understand this identity in the limit of $v \rightarrow 1- \epsilon$. 
    
    \[
    1 - v^2 = 1 - (1 - \epsilon)^2 = 2\epsilon - \epsilon^2 \approx 2\epsilon,
    \]
    so
    \[
    \gamma= \frac{1}{\sqrt{1 - v^2}} \approx \frac{1}{\sqrt{2\epsilon}}.
    \]
     which diverges as \(\epsilon \to 0^+\). The argument of \(f\) is
    \[
    \frac{(Z_0 + v Z_1)^2}{1 - v^2} = \gamma^2 (Z_0 + v Z_1)^2.
    \]
    Since \(v = 1-\epsilon\), let \(s = Z_0 + v Z_1 \approx Z_0 + Z_1 - \epsilon Z_1\). For small \(\epsilon\), the correction term \(\epsilon Z_1\) is negligible compared to the scale over which the expression peaks, so \(s \approx Z_0 + Z_1\). The left-hand side is then approximately \(\gamma f(\gamma^2 s^2)\).\\
    
    
    \textbf{for $\epsilon \rightarrow 0^+$ limit:}\\
    
    This expression \(\gamma f(\gamma^2 s^2)\) (with \(\gamma\) large) is a delta function in \(s\). it forms a narrow peak around \(s = 0\) (i.e., along the line \(Z_0 + Z_1 = 0\)) with width of order \(1/\gamma \approx \sqrt{2\epsilon}\). To derive the normalization, we consider integrating against a test function \(\psi(s)\):
    \[
    \int_{-\infty}^{\infty} \gamma f(\gamma^2 s^2) \psi(s) \, ds.
    \]
    Substitute \(k = \gamma s\), so \(ds = dk / \gamma\):
    \[
    \int_{-\infty}^{\infty} f(k^2) \psi(k / \gamma) \, dk.
    \]
    As \(\gamma \to \infty\) (i.e., \(\epsilon \to 0^+\)), \(\psi(k / \gamma) \to \psi(0)\), yielding
    \[
    \psi(0) \int_{-\infty}^{\infty} f(k^2) \, dk.
    \]
    This is precisely the action of \(\delta(s) \int_{-\infty}^{\infty} f(x^2) \, dx\) on \(\psi(s)\). The earlier approximation \(s \approx Z_0 + Z_1\) holds because the peak occurs where \(|s| \lesssim 1/\gamma \approx \sqrt{\epsilon}\), making the relative size of the correction \(\epsilon Z_1 / \sqrt{\epsilon} = \sqrt{\epsilon} Z_1 \to 0\) (assuming \(Z_1\) is not parametrically large).\\
    
    \textbf{Finite $\epsilon$:}\\
    
    Using the coordinate change \(s = Z_0 + v Z_1\), \(t = Z_1\) (with Jacobian determinant 1, so \(dZ_0 \, dZ_1 = ds \, dt\)) and \(v = 1 - \epsilon\), the integral becomes
    \[
    \int_{-\infty}^{\infty} \int_{-\infty}^{\infty} \frac{1}{\sqrt{1-v^2}} f\left( \frac{(Z_0 + v Z_1)^2}{1-v^2} \right) \psi(Z_0, Z_1) \, dZ_0 \, dZ_1 = \int_{-\infty}^{\infty} dt \int_{-\infty}^{\infty} ds \, \gamma f(\gamma^2 s^2) \psi(s - v t, t).
    \]
    Rescaling \(k = \gamma s\) (so \(ds = dk / \gamma\)) yields
    \[
    \int_{-\infty}^{\infty} dt \int_{-\infty}^{\infty} dk \, f(k^2) \psi\left( \frac{k}{\gamma} - v t, t \right).
    \]
    This can be rewritten as
    \[
    \int_{-\infty}^{\infty} dk \, f(k^2) \left[ \int_{-\infty}^{\infty} dt \, \psi\left( \frac{k}{\gamma} - v t, t \right) \right].
    \]
    
    The finite-\(\epsilon\) smearing appear in the test function: for each \(k\), \(\psi\) is evaluated at \((-t+k / \gamma + \epsilon t )\approx (-t+k \sqrt{2\epsilon}\)). Then we can approximate the test function at $-t$ by Taylor expansion.
    \begin{eqnarray}
        \psi(-t+k \sqrt{2\epsilon},t)=  \psi(-t,t)-k \sqrt{2\epsilon} \partial_t \psi(-t,t)+\frac{1}{2}k^2 \epsilon \partial^2_t \psi(-t,t)
    \end{eqnarray}
    Hence, the integral on the RHS becomes
    \begin{eqnarray}
    \int_{-\infty}^{\infty} dk \, f(k^2) \left[ \int_{-\infty}^{\infty} dt \, \psi\left( - t, t \right) \right]   && - \sqrt{2\epsilon} \int_{-\infty}^{\infty} dk \, f(k^2)k \left[ \int_{-\infty}^{\infty} dt \partial_t \psi(-t,t)\right]\nonumber\\
    &&+2\epsilon \int_{-\infty}^{\infty} dk \, f(k^2)k^2 \left[ \int_{-\infty}^{\infty} dt \partial^2_t \psi(-t,t)\right]
    \end{eqnarray}
    The implication of the first integral implies a delta function.
    The second integral vanishes because it is an odd function. The third term can also be evaluated. The effect of these terms is to smooth out the delta function. As \(\epsilon \to 0^+\) (\(\gamma \to \infty\)), these shifts vanish, recovering \(\int_{-\infty}^{\infty} dt \, \psi(-t, t)\)  thus the exact delta distribution form. For finite \(\epsilon\), this produces a weighted average over these nearby parallel lines, with weights \(f(k^2)\), effectively broadening the delta function but keeping the normalization integral same.\\

   \section{Shock wave energy and initial conditions:}
   \label{shockenergy}
    The shock wave parameter $p$ can be related to the initial conditions of the particle. We treat the black hole as a particle which starts from position $r=r_0$ (close to the origin) with radial velocity $\dot{r}_0$. At a late times, this particle will be accelerated to the speed of light and well approximated by a shockwave with energy $p$. We can thus relate the $p$ to the initial condition using the geodesic equation.\\

    A particle of mass $m$ in $dS_4$ has a Lagrangian
        \[
        \mathcal{L} = m \sqrt{-\big( -f(r)\dot{t}^2 + \frac{\dot{r}^2}{f(r)} \big)},
        \],
        where $f(r) = 1-\frac{r^2}{\ell^2}$. 
        The Killing vector of the static patch $\partial_t$ gives a conserved energy
        \[
        E = m f(r)\, \dot{t}.
        \]
        The normalization condition is $-f(r)\dot{t}^2 + \frac{\dot{r}^2}{f(r)} = -1$.
       It implies
        \[
        \dot{r}^2 = \frac{E^2}{m^2} - f(r).
        \]
        At the initial position $r = r_0$ with proper radial velocity $\dot{r}_0$, one finds
        \[
        E = m \sqrt{ \dot{r}_0^{\,2} + f(r_0)} =m  \sqrt{ \dot{r}_0^{\,2} + 1 - \frac{r_0^2}{L^2} }.
        \]
    At a late time, the energy \footnote{It is also the momentum $ p_v$.} of the particle is given by $p$. Hence, we can relate 
    \begin{eqnarray}
        p=m \sqrt{ \dot{r}_0^{\,2} + 1 - \frac{r_0^2}{L^2} }.
    \end{eqnarray}

\bibliographystyle{JHEP}
\bibliography{ref.bib}

\end{document}